\documentclass[
	a4paper, 
	10pt, 
	unnumberedsections, 
	twoside, 
]{LTJournalArticle}

\runninghead{} 

\footertext{\textit{RISC-V Summit Europe, Barcelona, 5-9th June 2023}} 

\title{Experiences of running an HPC RISC-V testbed} 

\author{
	Nick Brown\textsuperscript{1}\thanks{Corresponding author: \href{mailto:n.brown@epcc.ed.ac.uk}{\tt n.brown@epcc.ed.ac.uk}}, Maurice Jamieson\textsuperscript{1} and Joseph K. L. Lee\textsuperscript{1}
}

\date{\footnotesize\textsuperscript{\textbf{1}}EPCC, University of Edinburgh, Bayes Centre, 47 Potterrow, Edinburgh, United Kingdom}

\renewcommand{\maketitlehookd}{%
	\begin{abstract}
Funded by the UK ExCALIBUR H\&ES exascale programme, in early 2022 a RISC-V testbed for HPC was stood up to provide free access for scientific software developers to experiment with RISC-V for their workloads. Here we report on successes, challenges, and lessons learnt from this activity with a view to better understanding the suitability of RISC-V for HPC and important areas to focus RISC-V HPC community efforts upon.
	\end{abstract}
}

\begin{document}

\maketitle 


\section{Introduction}

The ExCALIBUR H\&ES RISC-V testbed has been operational for 12 months and our overarching aim has been to provide a service that HPC software developers will feel familiar with. This means providing access via a login node and the module environment enabling distinct versions of tooling such as compilers and libraries. RISC-V compute nodes are available via the Slurm scheduler using a shared filesystem throughout. At the time of writing, the testbed contains a diversity of nodes comprising six types of physical RISC-V CPUs and additionally soft-core RISC-V CPUs which are deployed via FPGAs to enable users to experiment run codes on cutting edge designs. 

Such a diversity of hardware and users makes this an interesting study in the current state of play of RISC-V for HPC workloads, and the purpose of this extended abstract is to summarise these insights and provide recommendations around how the ecosystem can be improved to help adoption. Interested readers can access the testbed at \emph{https://riscv.epcc.ed.ac.uk}.

\section{Testbed Experiences}
\subsection{Deceptively simple initial steps}

The RISC-V ecosystem is, certainly at first glance, impressive. It is no mean feat that supported, pretty much out of the box, is the ability to network different RISC-V boards and for these to run Linux which provides common tools such as NFS and Slurm. Furthermore, there are many libraries prebuilt for RISC-V, and this availability of common HPC libraries such as PETSC, FFTW, HDF5, and NETCDF is noteworthy.

The Message Passing Interface (MPI) is a ubiquitous standard for inter-node communication used in HPC providing, amongst other things, point to point and collective communications. The two main MPI implementations MPICH and OpenMPI are both available for RISC-V and consequently executing MPI codes across RISC-V compute nodes is no different to other technologies. Furthermore, binaries often run seamlessly across the distinct RISC-V CPUs in the testbed.

\subsection{Challenges}

Whilst the building blocks are all present in the RISC-V ecosystem for HPC, when delving deeper into setting up a service there have also been several challenges \cite{test-driving}. These fall into three major areas; \textbf{software tooling}, \textbf{OS kernels}, and \textbf{hardware availability}. Concerning software tooling, whilst there is an impressive set of libraries and compilers supporting RISC-V \cite{software-ecosystem}, support is lacking for common profiling tools, which are critical for HPC developers. Access to hardware performance counters is non-standard across the CPUs and whilst Perf is supported by some of the boards, often this requires rebuilding the stock kernel to enable and is not available on all RISC-V CPUs. It is our opinion that this should be a priority area of focus for HPC.

However, by far the largest challenge we found is in the vectorisation support provided by compilers. The main-line GCC compiler does not support RISC-V vectorisation because, whilst there was a v0.7.1 vectorisation branch, this was dropped. Indeed T-Head, the chip division of Alibaba, have provided their own version of GCC to explicitly support the v0.7.1 standard vectorisation of their RISC-V core, the XuanTie C906, which is used by the Allwinner D1 SoC. This in itself is a challenge because in recent months the download of this bespoke compiler GCC version has become inaccessible. 

By comparison, Clang supports RISC-V vectorisation at v1.0 only. However, there is a lack of hardware that currently supports v1.0, and far more implementing v0.7.1 for example built around the C906. This is challenging for HPC, where vectorisation is required for performance, and made worse by the lack of backwards compatibility between v1.0 and v0.7.1. This situation forced us to develop a tool which will modify the assembly code generated by Clang for v1.0 vectorisation and backport it to v0.7.1 \cite{roll-back}, enabling our users to compile via Clang and leverage vectorisation on testbed hardware. Furthermore, there are several ubiquitous HPC libraries that a large number of code rely upon, and whilst non-vectorised scalar versions of these are available for RISC-V, it is key that the community can optimise these by leveraging vectorisation. 

The Linux kernel that ships with the hardware can be overly restrictive and an example of this is with the DongshanNezhaSTU board, built around the Allwinner D1 SoC. It is not possible to compile new modules into the kernel on the device, for instance to support network file systems, due to the proprietary, protected, format of the bootloader. This means that the kernels need to be built externally (cross-compiled) whenever a new kernel is required and the bootloader must also be rebuilt at the same time. To allow kernel modules that enable the boards to be added to a cluster, such as network file system and device drivers, a new bootloader (enabling vectorisation at the hardware level) and new kernel based on Debian (exposing vectorisation to applications) must be built. To build a new Linux image, vendor specific patches must be applied to \emph{buildroot} and, in the case of the D1 use the T-Head specific GCC compiler version.

The third challenge we have faced is in the availability of hardware, especially those interesting for high performance workloads. Whilst there are numerous medium to long term efforts, it has been a challenge to source suitable machines for our testbed. This is one of the reasons why our testbed comprises a diverse mix of RISC-V boards, from the embedded Lichee RV Dock to the HiFive Unmatched and StarFive VisionFive V1 and V2. Furthermore this scarcity forced us to go down the route of providing soft-cores in the testbed, enabling us to provide cutting-edge features for users to experiment with, albeit at significantly reduced clock speed. Retrospectively, this was a useful choice as it enables experimentation with aspects likely present in future physical hardware.

\section{Benchmarking comparison}

\begin{figure}[h]
\centering
\includegraphics[width=\linewidth]{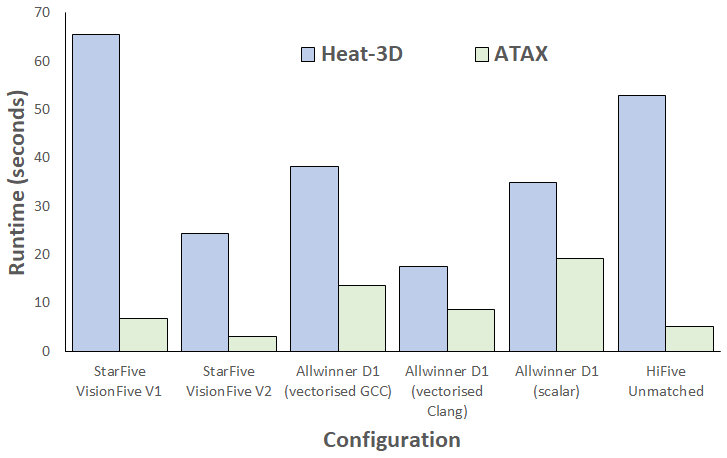}
\caption{Performance comparison of two Polybench kernels between testbed hardware}
\label{fig:performance}
\end{figure}

Users have been able to run a variety of applications and benchmarks on the RISC-V testbed, including atmospheric models such as MONC \cite{monc} and WRF, and quantum chemistry codes such as CP2K. These, in addition to extensive benchmarking, have resulted in numerous insights into the relative performance of hardware, many of which were surprising. Figure \ref{fig:performance} illustrates the performance of two benchmarks from Polybench \cite{yuki2014understanding}, compiled with Clang 16, \emph{Heat-3D} which is a stencil-based code calculating the Heat equation over a 3D data domain and \emph{ATAX} which is undertaking matrix transposition and vector multiplication. In this experiment we are executing over all the CPU cores; four in the case of the StarFive VisionFive V2 and HiFive Unmatched, two with the StarFive VisionFive V1 and one core for the Allwinner D1. There are two results for the Allwinner D1 with vectorisation, one using T-Head's GCC compiler and the second combining Clang with our v0.7.1 backport tool. It can be seen that vectorisation is beneficial on the Allwinner D1 when compiled with Clang as it produces the most efficient executable compared to GCC. Given that the Allwinner D1 costs around \$30, which is much less than the other boards, it is noteworthy that one vectorised core outperforms all other CPUs, comprising two or four cores, for the Heat-2D benchmark. 

\section{Conclusions \& recommendations}
Users of our RISC-V HPC testbed have been impressed that \emph{it just feels like any other HPC machine}, and whilst there are challenges standing up these technologies for HPC users, the key building blocks are present. We believe that RISC-V has a critical role to play in HPC, and as time progresses we will see RISC-V become ubiquitous in both specialist high performance CPUs and also accelerators. As a priority, the RISC-V HPC community should look to enhance support for vectorisation in libraries and compilers, as well as increasing the range of development tools, such as profilers that are available.

\section{Acknowledgement}
The authors would like to thank the ExCALIBUR H\&ES RISC-V testbed for access to compute resource and for funding this work. For the purpose of open access, the author has applied a Creative Commons Attribution (CC BY) licence to any Author Accepted Manuscript version arising from this submission.




\printbibliography 


\end{document}